# Simultaneous measurement of pressure-dependent bulk and interfacial thermal properties in thermal interface materials using square-pulsed source thermoreflectance

Tao Chen[1], Bingjia Xiao[1], Xin Qian[1], Puqing Jiang[1,*]

[1]*School of Energy and Power Engineering, Huazhong University of Science and Technology, Wuhan, Hubei 430074, China*

**ABSTRACT:** Thermal interface materials (TIMs) critically regulate heat dissipation from electronic chips to heat spreaders, yet their thermal conductivity (*k*), volumetric heat capacity (*C*), and interfacial thermal resistance (ITR) evolve with mechanical pressure and cannot be determined simultaneously using existing steady-state or transient techniques. As a result, the coupled roles of bulk compaction and interfacial contact in governing heat transport in TIM assemblies remain poorly resolved. Here, we present a square-pulsed source (SPS) thermoreflectance method that enables simultaneous determination of *k*, *C*, and ITR in TIM stacks under controlled mechanical loading. By spanning square-wave modulation frequencies from 1 Hz to 10 MHz, SPS probes a broad range of thermal penetration depths, enabling distinction between heat diffusion in the TIM bulk and interfacial heat transfer at the Al/TIM contact. Measurements on a thermally conductive gel, a thermal pad, and a high-vacuum grease during compression-unloading cycles reveal distinct pressure-dependent thermal transport mechanisms. The gel and pad exhibit increases in *k* and *C*, reduced ITR, and pronounced hysteresis, indicating coupled bulk densification and persistent interfacial conformity during loading cycles. In contrast, the grease exhibits bulk thermal properties that are weakly pressure-dependent within the measurement uncertainty, whereas its ITR remains strongly pressure-dependent. These results resolve the long-standing challenge of simultaneously quantifying bulk and interfacial thermal transport in mechanically loaded TIM assemblies, enabling experimentally constrained thermal management and reliability analysis in electronic packaging.

*Corresponding Author: jpq2021@hust.edu.cn (P.J.)

## 1. Introduction

As power densities continue to increase in microprocessors and artificial intelligence hardware [1], data centers [2], and power electronics [3], thermal management is increasingly limited not only by heat sinks and spreaders, but also by the thermal resistance along the heat-flow path from the heat-generating device to the external cooling structure [1, 3, 4]. In many practical packages, the dominant temperature drop can occur across the thermal interface between contacting solids rather than within the solids themselves. Thermal interface materials (TIMs) are therefore widely introduced to fill interfacial voids created by surface roughness, waviness, and non-planarity, thereby reducing the resistance to heat flow from the chip or module to the heat spreader, heat sink, or cold plate [1, 5]. In this sense, TIMs are not merely passive fillers; they are functional elements that directly regulate heat diffusion, heat storage, and contact heat transfer in the overall thermal process of electronic assemblies.

For a bonded TIM layer, thermal performance is governed by at least three quantities: the thermal conductivity k of the TIM bulk, the volumetric heat capacity C of the layer, and the interfacial thermal resistance (ITR) at the contacting surfaces. The first controls heat diffusion through the TIM, the second governs transient thermal storage within the layer, and the third reflects the efficiency of heat transmission across imperfect contact. While steady-state cooling analyses often emphasize only the overall thermal resistance, transient operation under pulsed power, load fluctuations, or thermal cycling also depends strongly on the thermal capacitance of the interface region [6]. Accordingly, meaningful evaluation of TIMs for modern packaging requires simultaneous knowledge of $k$, $C$, and ITR rather than any one parameter alone.

A further complication is that TIM properties are not always intrinsic constants under practical assembly conditions. Commercial TIMs are typically compressed during installation, and their internal structure and contact state can evolve with pressure. Previous work has shown that the overall thermal conductance of TIM joints may increase substantially with clamping pressure [7]. Thermal contact theory also indicates that asperity deformation and interfacial conformity during initial compression are only partially reversible, so that the same nominal pressure can correspond to different real contact areas during loading and unloading [8]. For soft TIMs, such as gels, greases, and compliant pads, compression may additionally alter void content, filler connectivity, local wetting, and bond-line morphology. These effects imply that the thermal response of the assembled interface can be both pressure-dependent and history-dependent. Such state dependence is directly relevant to packaging design because junction-to-case or junction-to-sink thermal impedance under service conditions is controlled by the actual compressed state of the TIM rather than by nominal vendor values measured under unrelated conditions.

The industrial standard ASTM D5470 [9] remains the most widely used method for TIM evaluation. However, as a steady-state one-dimensional method, it primarily yields an overall thermal resistance and does not determine the volumetric heat capacity. Moreover, separating intrinsic thermal conductivity from interfacial thermal resistance usually requires measurements at multiple bond-line thicknesses, which implicitly assumes that the TIM microstructure remains unchanged across the tested samples. This assumption becomes questionable for highly compliant TIMs, especially gels and greases that may undergo squeeze flow, phase redistribution, or pressure-induced structural rearrangement.

Transient macroscopic techniques such as laser flash analysis (LFA) can determine thermal diffusivity, i.e., $k/C$ [10], but cannot independently resolve $k$ and $C$ without auxiliary calorimetric input. In assembled TIM structures, LFA is also not well suited to separating bulk and interfacial contributions. Optical thermoreflectance techniques, particularly time-domain thermoreflectance (TDTR) [11-13] and frequency-domain thermoreflectance (FDTR) [6, 14], offer much higher spatial resolution and have become powerful tools for small-scale thermal-property measurements. Nevertheless, conventional TDTR is often mainly sensitive to thermal effusivity $\sqrt{kC}$ and ITR when applied to low-k materials [15, 16], rather than to $k$ and $C$ separately. The accessible frequency range is also limited in practice, because conventional TDTR is usually operated above about 0.1 MHz, where lock-in detection remains robust. Low-frequency thermoreflectance approaches have recently extended sensitivity to $k$ and $C$ in TIMs [6, 17], but the low-frequency range that benefits bulk-property decoupling is generally insufficient to probe interfacial thermal resistance strongly. As a result, simultaneous determination of $k$, $C$, and ITR in soft TIMs remains challenging.

Another important gap concerns realistic mechanical states. Existing thermoreflectance studies of TIMs under pressure are still limited, and $C$ is often treated as pressure-invariant during inversion. For soft materials, however, volumetric heat capacity may change with compression because the volume fraction of air, polymer-rich phase, or filler-rich phase evolves. If $C$ is fixed a priori when it is actually state-dependent, the extracted $k$ and ITR can be biased. More broadly, the lack of an experimental framework that can track the path-dependent evolution of bulk and interfacial thermal properties during loading-unloading cycles has hindered mechanistic understanding of TIM behavior under realistic assembly conditions.

The square-pulsed source (SPS) thermoreflectance technique provides an opportunity to address this problem. In SPS, the pump beam is modulated by a 50% duty-cycle square wave over an exceptionally broad frequency window from 1 Hz to 10 MHz [15, 18], and the time-resolved thermoreflectance response is interpreted with a heat-transfer model. This broad range allows the thermal penetration depth to vary over several orders of magnitude, thereby providing access to distinct timescales of heat storage, heat diffusion, and interfacial heat transfer within a single experimental framework. SPS has previously been shown to simultaneously determine thermal conductivity and volumetric heat capacity across diverse materials [15, 19], and it has also been extended to multilayer structures and convective heat-transfer measurements [18, 20]. However, its capability for pressure-controlled TIM assemblies, where bulk and interfacial properties evolve together, has not yet been demonstrated.

Against this background, the present study addresses three knowledge gaps: the lack of a single transient method that simultaneously resolves $k$, $C$, and ITR in TIM assemblies; the limited understanding of how these parameters evolve with pressure and loading history; and the insufficient linkage between TIM property characterization and the thermal processes that govern electronic packaging performance. To address these gaps, we combine SPS thermoreflectance with a pressure-controlled sample platform and examine three representative TIM classes: a thermally conductive gel, a thermal pad, and a high-vacuum grease. The applied pressure range of 0.05-0.18 MPa falls within the regime relevant to practical engineering applications. Section 2 describes the SPS principle, the pressure-controlled measurement configuration, the tested materials, and the uncertainty framework. Section 3 then presents representative signal decoupling and the pressure-dependent thermal behavior of the three TIMs, followed by physics-based interpretation of their bulk-state and interfacial evolution. By doing so, this work aims to establish a design-relevant route for characterizing state-dependent thermal transport in TIM assemblies used in practical thermal-management systems.

## 2. Materials and methods

### *2.1 Square-Pulsed Source (SPS) technique and pressure-controlled setup*

The SPS method is a non-contact pump-probe thermoreflectance technique based on the approximately linear dependence of probe reflectivity on surface temperature. A schematic of the experimental setup is shown in Fig. 1(a). The pump laser is square-wave modulated by the lock-in amplifier to periodically heat the sample surface, while the reflected probe beam is collected by a photodetector. The periodic waveform analyzer (PWA) module of the lock-in amplifier records the time-dependent voltage response over one modulation period. The corresponding temporal profile of the pump intensity is illustrated in Fig. 1(b).

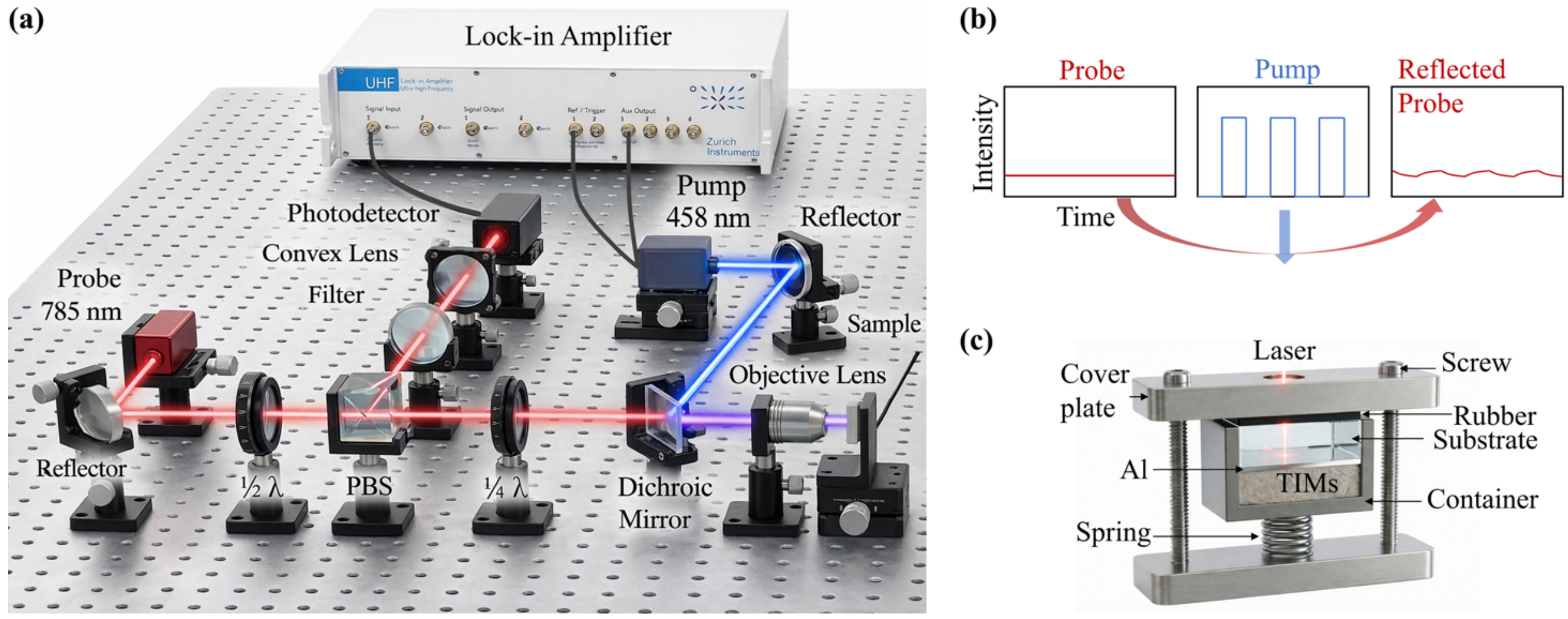

FIG.1 Experimental configuration of the square-pulsed source (SPS) thermoreflectance technique under compression. (a) Optical setup of the SPS pump-probe system. (b) Schematic temporal profiles of the probe, pump, and reflected probe signals within one modulation period. (c) Pressure-controlled sample holder used for TIM measurements, in which an Al-coated transparent substrate is pressed against the TIM by a spring-loaded fixture.

A key advantage of SPS is its exceptionally broad modulation-frequency range, spanning from 1 Hz to 10 MHz. This range enables access to thermal responses over widely different penetration depths and timescales, which is essential for separating bulk and interfacial contributions within the same experiment. In the present TIM problem, low frequencies enhance sensitivity to heat diffusion and heat storage in the TIM bulk, whereas high frequencies strengthen sensitivity to interfacial heat transfer at the Al/TIM contact. Details of the heat-transfer model are provided in Supplementary Information Section S1.

To enable measurements under compression, SPS is integrated with a pressure-controlled platform, as illustrated in Fig. 1(c). An approximately 100 nm thick Al transducer layer is first deposited onto a transparent substrate. The coated substrate is then flipped and brought into contact with the TIM under a controlled load. The applied force is obtained from spring compression using Hooke's law, and the nominal pressure is calculated by dividing the force by the cross-sectional area of the TIM container. Both pump and probe beams pass through the transparent substrate and are focused at the substrate/Al interface [21]. The measured transient thermoreflectance signal is then fitted with an appropriate heat-conduction model to extract the TIM thermal conductivity, volumetric heat capacity, and interfacial thermal resistance.

*2.2 Test materials and measurement protocol*

Glass was used as the transparent substrate because its relatively low thermal conductivity directs a larger fraction of the heat flux toward the TIM, thereby reducing uncertainty in the extracted TIM properties. The Al transducer layer was approximately 100 nm thick and was deposited by electron-beam evaporation. The tested TIMs included a thermally conductive gel (RS382, provided by Shenzhen Thermal Acoustic Intelligent Co., Ltd.), a thermal pad (Laird HD90000), and a high-vacuum grease purchased from Dow Corning. In the present study, the thicknesses of the thermally conductive gel and thermal pad were much greater than the thermal penetration depth in the TIM, and their lateral dimensions were also much larger than the laser spot radius. They could therefore be treated as bulk materials, such that neither their thickness variation nor the interfacial thermal resistance between the TIM and the container needed to be considered. Although part of the grease was squeezed out of the container during compression, its thickness after loading still remained greater than 1 mm, which was much larger than the thermal penetration depth in the TIM.

For each TIM, measurements were performed during controlled compression-unloading cycles. At each pressure level, SPS signals were acquired at multiple modulation frequencies and fitted simultaneously. This protocol was designed to capture not only the pressure dependence of the extracted parameters, but also any hysteresis associated with irreversible or partially reversible evolution of the bulk TIM state and the Al/TIM contact state. Such a loading-history perspective is important because practical electronic packages may experience initial assembly, rework, stress relaxation, and thermal cycling, all of which can alter the real thermal state of the interface.

*2.3 Sensitivity analysis and uncertainty quantification*

Sensitivity analysis was used to evaluate the extent to which a parameter can be reliably determined from the measured signal. The sensitivity coefficient is defined as

$$S_{\xi} = \frac{\partial \ln A_{\mathrm{norm}}}{\partial \ln \xi} \tag{1}$$

where $\xi$ denotes a parameter of interest and $A_{\mathrm{norm}}$ is the normalized amplitude signal. Physically, $S_{\xi}$ indicates the percentage change in signal caused by a 1% change in parameter $\xi$. In general, $|S_\xi| > 0.2$ indicates strong sensitivity, whereas $|S_\xi| < 0.05$ suggests weak sensitivity.

To quantify the uncertainty of the extracted parameters, we adopted a full error-propagation framework that accounts not only for uncertainties in known input parameters, but also for experimental noise and fitting residuals. This approach, originally introduced by Yang et al. for FDTR uncertainty analysis [22], is particularly suitable for multiparameter inverse problems.

In the fitting procedure, $M$ sets of experimental signals are regressed simultaneously. The objective function is defined as the product of the mean squared deviations (MSDs) of all signal sets:

$$J = \prod_{j=1}^{M} \mathrm{MSD}_j \tag{2}$$

where $\mathrm{MSD}_j$ denotes the MSD between the $j$-th measured signal and its corresponding model prediction. The optimal solution is obtained by minimizing $J$.

Near the optimum, the uncertainty of the fitted parameters is described by their covariance matrix,

$$\mathrm{Var}[\hat{X}_U] = \begin{bmatrix} \sigma_1^2 & \mathrm{cov}_{12} & \cdots \\ \mathrm{cov}_{21} & \sigma_2^2 & \cdots \\ \vdots & \vdots & \ddots \end{bmatrix} \tag{3}$$

where the diagonal terms are the parameter variances and the off-diagonal terms quantify parameter covariance, i.e., parameter coupling. In this work, the uncertainty of each extracted parameter is reported as twice its standard deviation ($2\sigma$). The detailed derivation is provided in Section S2 of the Supplementary Information.

## 3. Results and discussion

### *3.1 Representative measurement and parameter decoupling*

We first adopted a differential measurement strategy to characterize the reference glass/Al/air structure. A laser spot radius of 12.2 $\mu$m was used. This spot size was selected as a practical compromise: a significantly smaller spot would increase sensitivity to the transducer properties, whereas a much larger spot would reduce the signal-to-noise ratio. Three pump modulation frequencies spanning different timescales, namely 1 MHz, 100 kHz, and 4 kHz, were employed. By simultaneously fitting the signals measured at these three frequencies, the thermal conductivity and volumetric heat capacity of the glass substrate, as well as the thermal conductivity of the Al film, were determined [21].

We then considered a representative case for the thermally conductive gel at 0.16 MPa to illustrate how SPS decouples the thermal conductivity, volumetric heat capacity, and interfacial thermal resistance of the Al/TIM system. The measurements were performed using the same 12.2 $\mu$m laser spot radius and the same three modulation frequencies. The corresponding normalized amplitude signals are shown in Fig. 2(a-c), and the associated sensitivity coefficients are plotted in Fig. 2(d-f).

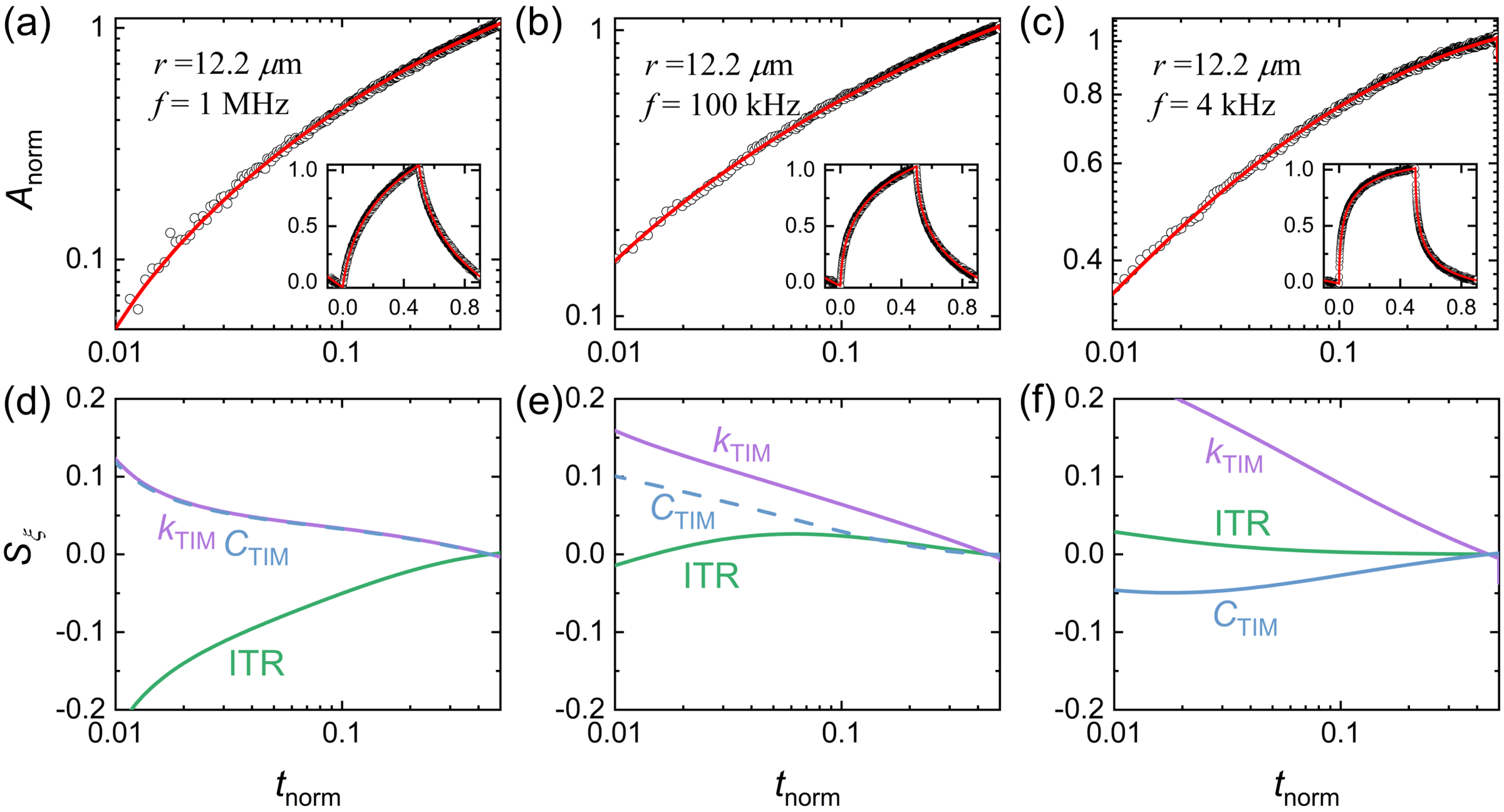

FIG.2 Representative SPS measurement and sensitivity analysis for the thermally conductive gel at 0.16 MPa. (a-c) Normalized amplitude signals measured with a laser spot radius of 12.2 $\mu$m at modulation frequencies of 1 MHz, 100 kHz, and 4 kHz, respectively. Symbols are experimental data and solid lines are model fits; the insets show the corresponding full-period transient responses. (d-f) Sensitivity coefficients to the TIM thermal conductivity $k_{\mathrm{TIM}}$, volumetric heat capacity $C_{\mathrm{TIM}}$, and ITR.

At 1 MHz, the sensitivities to $k_{\mathrm{TIM}}$ and $C_{\mathrm{TIM}}$ nearly overlap, while the sensitivity to ITR is particularly strong. This indicates that under high-frequency heating the signal is governed mainly by the thermal effusivity $\sqrt{kC}$ together with the interfacial thermal resistance. At 100 kHz, the sensitivities to $k_{\mathrm{TIM}}$ and $C_{\mathrm{TIM}}$ begin to separate, although some coupling remains, while the sensitivity to ITR decreases. At 4 kHz, the sensitivities to $k_{\mathrm{TIM}}$ and $C_{\mathrm{TIM}}$ exhibit opposite signs, indicating effective decoupling of these two quantities. Accordingly, simultaneous fitting of the three signals provides sufficient independent information to determine all three unknown parameters, yielding $k_{\mathrm{TIM}} = 2.24\ \mathrm{W/(m \cdot K)}$, $C_{\mathrm{TIM}} = 0.36\ \mathrm{MJ/(m^3 \cdot K)}$, and $\mathrm{ITR} = 0.77\ \mathrm{mm^2 \cdot K/W}$.

This result is important from a thermal-process perspective. It shows that SPS can access, within one experimental framework, the three mechanisms that govern heat flow across a TIM joint: heat storage in the compliant layer, heat diffusion through the TIM bulk, and contact heat transfer across the interface. That capability forms the basis for the pressure-dependent analysis below.

### *3.2 Thermally conductive gel: coupled bulk densification and interface evolution*

Figure 3 summarizes the pressure-dependent thermal behavior of the RS382 thermally conductive gel during loading and unloading. Three clear trends are observed during loading: thermal conductivity increases with pressure, volumetric heat capacity also increases, and interfacial thermal resistance decreases. Specifically, $k$ rises from 2.24 W/(m·K) at 0.16 MPa to 4.4 W/(m·K) at 1.85 MPa, $C$ increases from 0.36 to 0.68 MJ/($\mathrm{m^3}$·K), and ITR decreases continuously over the same pressure range. These results indicate that compression enhances both bulk and interfacial heat transport.

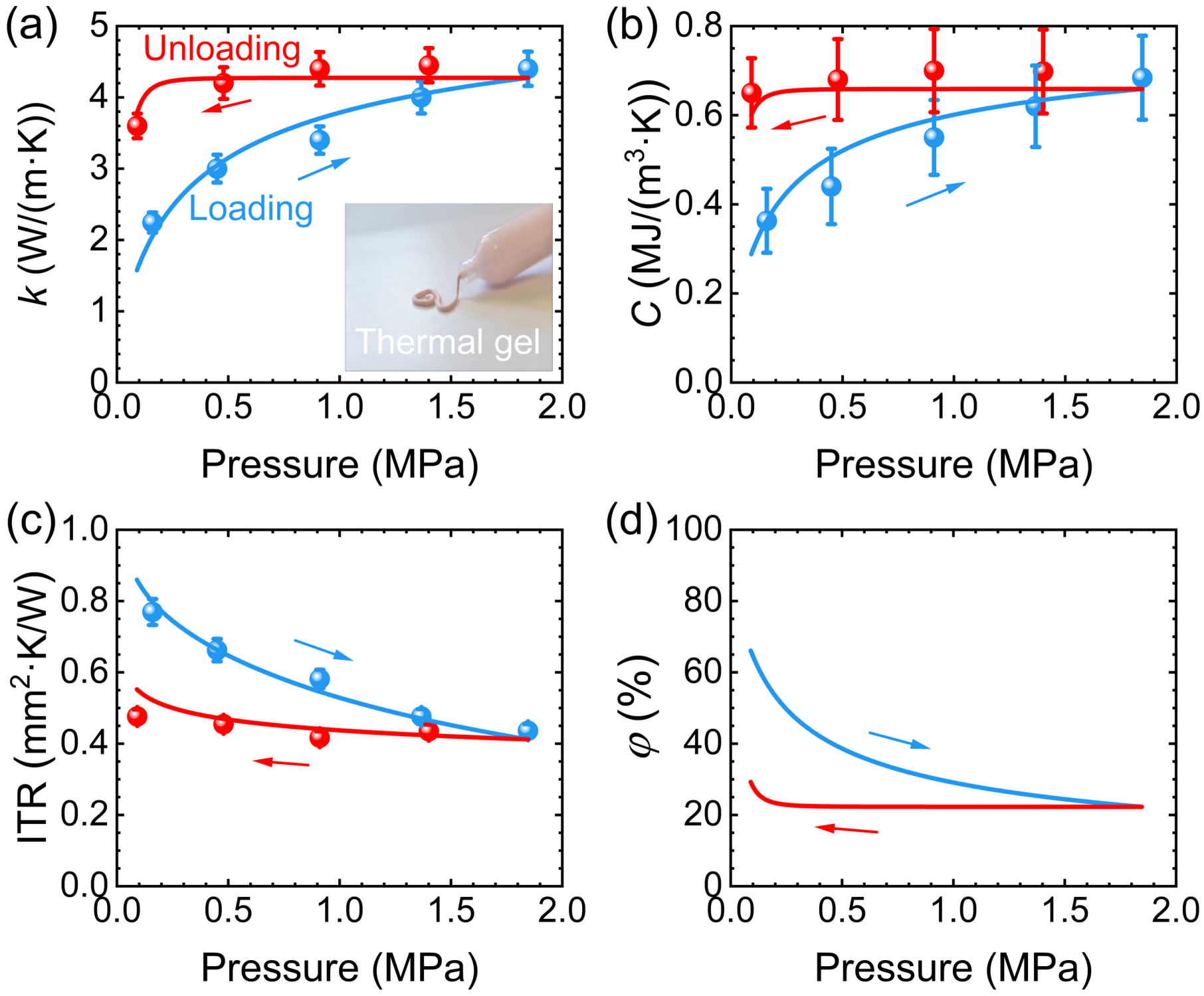

FIG.3 Pressure-dependent thermal properties of the thermally conductive gel during loading and unloading. (a) $k$. (b) $C$. (c) Al-TIM ITR. Symbols with error bars are SPS measurements and solid lines are fits of the pressure-dependent porous-medium/contact model. (d) Inferred porosity evolution $\varphi$. Blue denotes loading and red denotes unloading.

For comparison, the same thermally conductive gel was also measured by a steady-state method, yielding an apparent thermal conductivity of approximately 8 W/(m·K), which is substantially higher than the thermal conductivity obtained from SPS. This difference is physically reasonable and does not indicate inconsistency between the two methods. As pointed out in previous work [23], transient bulk measurements and ASTM D5470 steady-state measurements probe different thermal states of TIMs. The former is closer to the as-dispensed or bulk-like response of the material, whereas the latter reflects heat conduction through a compressed thin bondline under application conditions. In filler-loaded TIMs, reducing the bondline thickness to a scale comparable to the filler size can break the assumption of statistical homogeneity, enhance through-plane anisotropy, and promote more direct heat conduction through filler-filler contacts. As a result, the apparent thermal conductivity obtained by the steady-state method can be significantly higher than the bulk thermal conductivity determined by SPS. As a further check on the magnitude of the SPS result, the present low-pressure thermal conductivity is also close to literature values of about 3 W/(m·K) reported for commercial gel-type TIMs under bulk-like transient measurements [23], indicating that the SPS result is within a reasonable range.

More importantly, all three properties exhibit pronounced hysteresis. At the same nominal pressure, the unloading branch shows a higher $k$, a higher $C$, and a lower ITR than the loading branch. For example, at 0.48 MPa, the unloading thermal conductivity remains close to 4.2 W/(m·K), whereas the loading value is only about 3.0 W/(m·K). Similar differences are observed in $C$ and ITR. This behavior shows that the thermal state of the gel is governed not only by the instantaneous pressure, but also by the prior compression history.

To interpret these trends, we model the gel as a pressure-dependent compressible porous medium with evolving interfacial contact. This model is not chosen empirically, but follows directly from the physical characteristics of the gel, which consists of a continuous soft matrix and a dispersed low-conductivity gas/void phase. Compression reduces the fraction of the low-conductivity phase and simultaneously improves interfacial conformity. During loading, the porosity is described as

$$\varphi_L(P) = \varphi_{\mathrm{res},L} + \varphi_{\mathrm{comp},L}\left(\frac{P_{\mathrm{ref}}+P_{\mathrm{atm}}}{P+P_{\mathrm{atm}}}\right)^{1/n_L} \tag{4}$$

where $\varphi_{\mathrm{res},L}$ is the residual porosity at high pressure, $\varphi_{\mathrm{comp},L}$ is the compressible porosity component, $P_{\mathrm{atm}}$ is atmospheric pressure, $P_{\mathrm{ref}}$ is the reference pressure at the initial loading state, and $n_L$ is an effective compression exponent. Equation (4) is derived in this work by assuming that a compressible fraction of the porosity originates from entrapped gas pockets at the TIM-solid interface, which are compressed under pressure [24-26]. The pressure dependence follows a polytropic relation $P_{\text{abs}}V^n = \text{const}$ [27], while a residual porosity term accounts for non-collapsible microvoids.

During unloading, the porosity follows a different branch,

$$\varphi_U(P) = \varphi_{\text{res},U} + \left[\varphi_L(P_{\max}) - \varphi_{\text{res},U}\right]\left(\frac{P_{\max}+P_{\text{atm}}}{P+P_{\text{atm}}}\right)^{1/n_U} \tag{5}$$

where $\varphi_{\text{res},U}$ is the residual porosity during unloading, $n_U$ is an effective recovery exponent, and $P_{\max}$ is the maximum applied pressure.

The volumetric heat capacity is described by a rule of mixtures [28],

$$C = (1-\varphi)C_s + \varphi C_{\text{air}} \tag{6}$$

where $C_s$ is the volumetric heat capacity of the solid-rich skeleton and $C_{\text{air}}$ is that of the pore phase. The effective thermal conductivity is represented by a Maxwell porous-medium model [29],

$$k_{\text{eff}} = k_s \frac{2k_s+k_{\text{air}}-2(k_s-k_{\text{air}})\varphi}{2k_s+k_{\text{air}}+(k_s-k_{\text{air}})\varphi} \tag{7}$$

where $k_s$ and $k_{\text{air}}$ are the thermal conductivities of the solid skeleton and pore phase, respectively. The interfacial response is described by a composite conductance model [30-32],

$$\text{ITR} = 1/(Ak^*P^{0.95} + BP^{0.097}) \tag{8}$$

where $k^*$ is the harmonic mean of the thermal conductivity of Al and that of the TIM, and $A$ and $B$ are fitting parameters.

The fitted curves shown in Fig. 3(a-c) agree well with the experimental data, supporting the applicability of this framework. The inferred porosity evolution in Fig. 3(d) provides a physically consistent explanation for the measured hysteresis. During loading, porosity decreases rapidly with pressure, indicating collapse of void space and bulk densification. Because the pore phase has much lower thermal conductivity and heat capacity than the solid-rich matrix, this densification naturally increases both $k$ and $C$. During unloading, however, the porosity remains lower than that on the loading branch at the same pressure, indicating incomplete structural recovery. As a result, the gel retains a denser state after compression, which explains the higher unloading $k$ and $C$.

The lower unloading ITR can be interpreted in a similar way. Once compressed, the gel conforms more closely to the Al surface, and the real contact area established at high pressure is only partially lost during unloading. In addition, viscoelastic relaxation and possible irreversible deformation at local contact spots may preserve a more favorable thermal contact state. Consequently, the interface remains thermally better connected during unloading than during first loading at the same nominal pressure.

These results have direct implications for practical thermal management. In many package-level analyses, the TIM is assigned a single conductivity and the heat capacity is treated as fixed or ignored. The present measurements show that for a compliant gel, $k$, $C$, and ITR can all depend on pressure and loading history. Neglecting this state dependence may therefore misrepresent transient thermal impedance and temperature excursion in applications with nontrivial assembly pressure or stress relaxation.

*3.3 Thermal pad: compressed-state evolution with strong hysteresis*

We next examine whether the pressure- and history-dependent behavior observed in the gel also occurs in a more structured compliant TIM. Figure 4 summarizes the thermal properties of the commercial thermal pad during loading and unloading. As in the gel case, $k$, $C$, and ITR all vary strongly with pressure. However, the underlying physical interpretation is different, because the pad is better regarded as a highly compliant filled composite than as a simple porous gel.

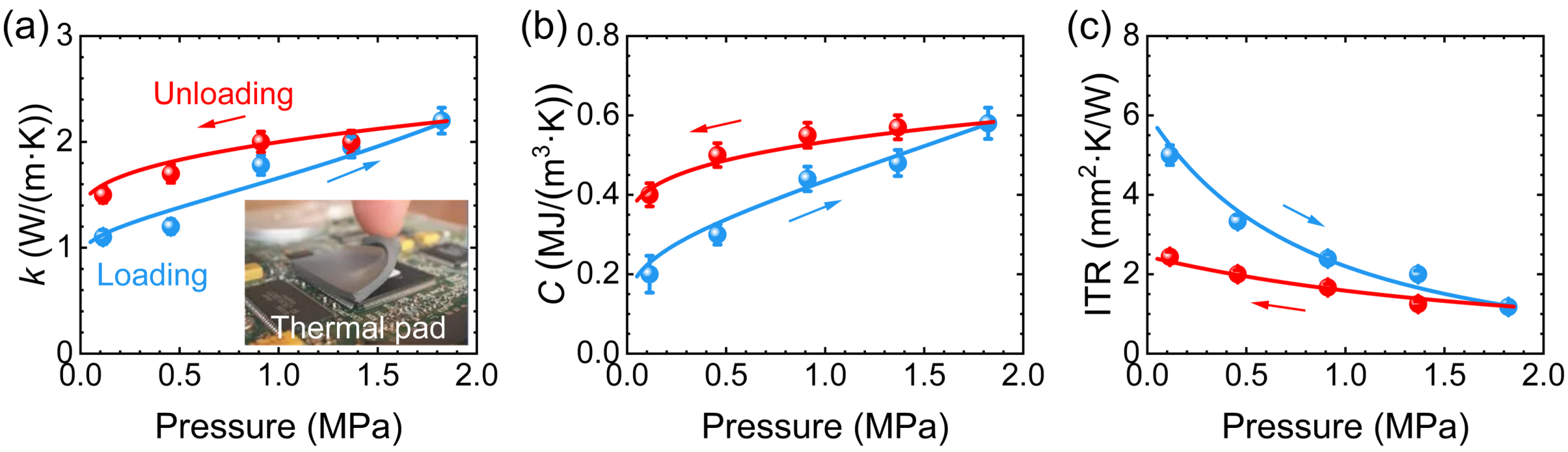

FIG.4 Pressure-dependent thermal properties of the thermal pad during loading and unloading. (a) $k$. (b) $C$. (c) Al-TIM ITR. Symbols with error bars are SPS measurements and solid lines are fits of the reduced compressed-state/contact-conductance model. Blue denotes loading and red denotes unloading.

During loading, the thermal conductivity increases from 1.1 W/(m·K) at 0.11 MPa to 2.2 W/(m·K) at 1.82 MPa. Over the same pressure range, the volumetric heat capacity rises from 0.20 to 0.58 MJ/($m^3$·K), while ITR decreases markedly from 5.0 to 1.2 $mm^2$·K/W. These trends indicate that compression improves both bulk thermal transport and the interfacial heat-transfer condition.

A clear hysteresis is again observed. At a given pressure, the unloading branch exhibits a higher $k$, a higher $C$, and a lower ITR than the loading branch, especially at low pressure. For example, at 0.11 MPa, the unloading thermal conductivity remains close to 1.5 W/(m·K), compared with only about 1.1 W/(m·K) during first loading. Similar differences are observed in $C$ and ITR. At higher pressures, the loading and unloading branches gradually converge. This indicates that the thermal state of the pad is also path-dependent, although the relevant internal state variable differs from that of the gel.

To interpret these results, we adopt a reduced hybrid model based on the pressure-dependent compressed state of the pad. Rather than introducing an explicit porosity-based effective-medium description, we characterize the mechanical state using a Tait-type pressure-specific-volume relation [33]. The loading and unloading branches are written as

$$v_L(P) = v_{0,L}\left[1 - C_T \ln\left(1+\frac{P}{B_L}\right)\right] \tag{9}$$

and

$$v_U(P) = v_{0,U}\left[1 - C_T \ln\left(1+\frac{P}{B_U}\right)\right] \tag{10}$$

where $v$ is the specific volume, $C_T$ is the Tait constant, and $B_L$ and $B_U$ are fitting parameters, with continuity imposed at the maximum pressure. The initial specific volume in the loading branch, $v_{0,L}$, is fixed at $2.86 \times 10^{-4}$ m$^3$/kg according to the value provided by the manufacturer.

Within this framework, the volumetric heat capacity is expressed as

$$C(P) = \frac{c_{p,\mathrm{eff}}}{v(P)} \tag{11}$$

where $c_{p,\mathrm{eff}}$ is the effective mass-specific heat capacity. Compression reduces the specific volume and therefore increases the volumetric heat capacity. This explains both the monotonic rise of $C$ during loading and its persistently higher value during unloading.

The thermal conductivity is described by a reduced state-dependent relation [34],

$$\ln\ k(P) = a + \frac{b}{v(P)} \tag{12}$$

where $a$ and $b$ are effective fitting parameters. This expression captures the experimentally observed increase in $k$ as the pad is compressed into a denser state. Physically, compression is expected to strengthen the internal heat-conduction pathways within the filled composite by reducing internal separation and improving network connectivity.

The interfacial response is modeled within a contact-conductance framework based on the Cooper-Mikic-Yovanovich (CMY) model [30]. The corresponding interfacial thermal resistance is written as

$$\mathrm{ITR}(P) = 1/(Ak^*(P)P^{0.95} + G_g) \quad (13)$$

where $k^*(P)$ is the harmonic mean of the thermal conductivity of Al and that of the TIM, $A$ is an effective contact parameter, and $G_g$ represents the gap-related conductance contribution. Separate loading and unloading branches are used while maintaining continuity at the maximum pressure.

The fitted curves shown in Fig. 4(a-c) reproduce the measured trends well. The inferred compressed-state evolution provides a consistent physical explanation for the hysteresis: during loading, the pad becomes progressively compacted, which raises both $C$ and $k$ and lowers ITR; during unloading, the material does not fully recover its original state, so the specific volume remains lower than that on the loading branch at the same nominal pressure. This residual compression set leads directly to higher unloading $k$, higher unloading $C$, and lower unloading ITR.

Although the overall hysteresis direction resembles that of the gel, the physical picture is different. In the gel, the behavior can be interpreted primarily in terms of porosity evolution in a compressible porous medium. In the thermal pad, the more appropriate description is a pressure-dependent compressed bulk state together with enhanced interfacial conformity. The key implication, however, is the same: the thermal properties of the pad cannot be treated as fixed material constants under realistic assembly conditions. Instead, they are state-dependent quantities governed by pressure and mechanical history. For package designers, this means that the thermal response of pad-based joints under startup, re-clamping, or stress relaxation may differ substantially from that predicted using a single nominal property set.

### *3.4 High-vacuum grease: interface-dominated pressure response*

A distinct behavior is observed for the high-vacuum grease, as summarized in Fig. 5. Unlike the thermally conductive gel and the thermal pad, the grease shows little measurable pressure dependence in its bulk thermal conductivity and volumetric heat capacity over the investigated pressure range. As shown in Fig. 5(a), the extracted thermal conductivity remains approximately constant at about 0.15 W/(m·K), which is in reasonable agreement with the manufacturer-reported value of 0.19 W/(m·K). Similarly, Fig. 5(b) shows that the volumetric heat capacity stays within a narrow range of roughly 0.65-0.70 MJ/($m^3$·K), with variations comparable to the experimental uncertainty.

This behavior differs fundamentally from that of the gel and pad. In those materials, compression alters the internal bulk state, leading to systematic increases in both $k$ and $C$. For the grease, no such trend is observed, suggesting that within the present pressure range the intrinsic bulk thermophysical properties are largely insensitive to compression. Instead, the dominant effect of pressure is interfacial.

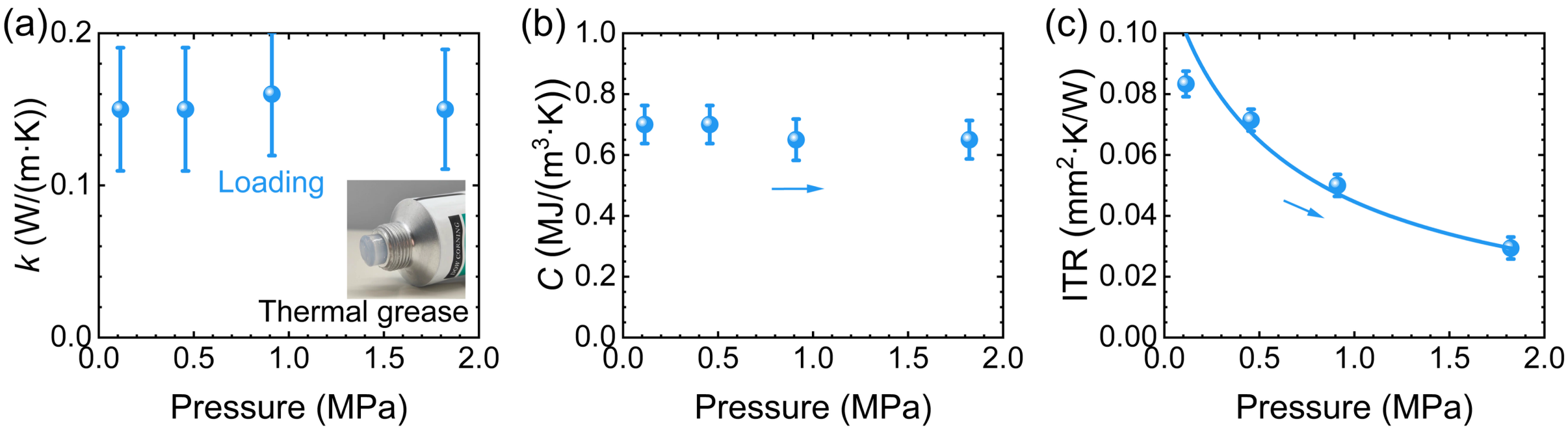

FIG.5 Pressure-dependent thermal properties of the high-vacuum grease. (a) $k$. (b) $C$. (c) Al-TIM ITR. Symbols with error bars are SPS measurements and solid lines are fitted trends.

As shown in Fig. 5(c), the Al-TIM interfacial thermal resistance decreases significantly with increasing pressure, from 0.08 $mm^2$·K/W at low pressure to 0.03 $mm^2$·K/W at the highest pressure. This strong pressure dependence indicates that thermal improvement in the grease is governed mainly by enhanced wetting and gap filling at the interface. With increasing load, the grease is more effectively squeezed into surface asperities and residual voids, thereby reducing local separation and increasing the effective heat-transfer area across the interface.

To quantify this behavior, the ITR is fitted using the same composite interfacial conductance framework introduced above. Because the bulk thermal conductivity of the grease remains nearly constant, the pressure dependence of ITR can be attributed primarily to evolving interfacial conformity rather than to changes in bulk transport. The fitted curve captures the experimental data well, confirming that the main role of pressure in this material is to improve the interface.

These observations have an important physical implication. For grease-type TIMs, the bulk material behaves approximately as an invariant thermal medium under the present loading conditions, whereas the interface remains highly pressure-sensitive. In other words, the thermal response is governed predominantly by how effectively the grease wets and bridges the contacting solids, rather than by compression-induced changes in its intrinsic bulk state. This behavior is consistent with the intended function of greases as highly conformable interstitial fillers.

Taken together, the results in Fig. 5 complement those of Figs. 3 and 4 by showing that different classes of TIMs respond to pressure through fundamentally different mechanisms. Whereas gels and pads exhibit coupled bulk-state and interface evolution, grease responds mainly through interface evolution. This contrast underscores the importance of separating bulk and interfacial contributions when evaluating TIMs under realistic assembly conditions. It also indicates that, for grease-based interfaces, engineering optimization should focus primarily on wetting, spreading, and contact formation, while for gels and pads both bulk-state evolution and contact evolution must be considered.

## 4. CONCLUSIONS

This work presents a square-pulsed source (SPS) thermoreflectance method for the simultaneous measurement of thermal conductivity, volumetric heat capacity, and interfacial thermal resistance in thermal interface material assemblies under compressive loading. The originality of the study lies in resolving, within a single transient measurement framework, the path-dependent evolution of both bulk and interfacial thermal properties under realistic assembly conditions. By combining a broad modulation-frequency range with time-resolved thermoreflectance detection and inverse analysis, the method separates heat storage, heat diffusion, and interfacial heat transfer without requiring the heat capacity to be prescribed a priori.

Measurements on a thermally conductive gel, a thermal pad, and a high-vacuum grease reveal distinct pressure-dependent thermal processes. For the gel and pad, both bulk properties and interfacial thermal resistance evolve significantly with pressure and show clear loading-unloading hysteresis, indicating persistent changes in bulk state and interfacial conformity. For the grease, in contrast, the bulk thermal conductivity and volumetric heat capacity remain nearly unchanged, while the interfacial thermal resistance decreases markedly with pressure, indicating an interface-dominated response.

These findings show that TIM properties under practical assembly conditions should not always be treated as fixed constants, but rather as state-dependent quantities controlled by pressure and mechanical history. From an application perspective, accurate prediction of transient thermal impedance and thermal reliability in electronic packaging requires simultaneous consideration of $k$, $C$, and ITR under the actual compressed state of the TIM. More importantly, the observed hysteresis provides direct guidance for TIM assembly and preload design. For gel- and pad-based interfaces, a controlled pre-compression step can help preserve a denser bulk state and better interfacial conformity, thereby improving long-term thermal performance and reducing interfacial thermal resistance during subsequent operation. For pad-based interfaces in particular, a sufficiently high initial assembly pressure is needed to overcome low-pressure contact limitations and mitigate hysteresis-related performance loss.

For grease-based interfaces, by contrast, increasing preload mainly improves interfacial heat transfer. The SPS approach developed here therefore provides both a measurement tool and a physically informative framework for the thermal design and assessment of TIM-based heat-flow paths in advanced electronic systems.

## ACKNOWLEDGMENT

X.Q. acknowledges the funding support from National Key R&D program (2022YFA1203100). P.J. acknowledges support from the National Natural Science Foundation of China (NSFC) through Grant No. 52376058.

## DATA AVAILABILITY

The data that support the findings of this study are available from the corresponding author upon reasonable request.